%% file: main.tex
\documentclass{llncs}
\usepackage[utf8]{inputenc}

\usepackage{makeidx}  
\usepackage{url}
\usepackage{xspace} 
\newcommand{\PC}{\textsc{Prismacloud}\xspace}
\newcommand{\todo}[1]%
  {\marginpar{\baselineskip0ex\rule{2,5cm}{0.5pt}\\[0ex]{\tiny\textsf{#1}}}}

\usepackage{tikz}
\usetikzlibrary{calc}
\usepackage{atbegshi}
\AtBeginShipoutNext{%
    \AtBeginShipoutUpperLeft{%
        \begin{tikzpicture}[remember picture,overlay]%
            \tikzstyle{every node}=[font=\sf\normalsize]%
            \node [color=lightgray, text width=15cm, text centered]%
                at ($ (current page.north)+(0,-1) $) {
                4th Cyber Security and Privacy EU Forum, CSP Forum 2015, Revised Selected Papers, to Appear.\\
                Author’s accepted version.\\
            };%
        \end{tikzpicture}
    }
}

\begin{document}
\mainmatter              
\title{Towards a New Paradigm for Privacy and Security in Cloud Services}

\titlerunning{PRISMACLOUD}  
%


\author{Thomas Lor{\"u}nser\inst{1}
\and Charles Bastos Rodriguez\inst{2} 
\and Denise Demirel\inst{3} 
\and Simone Fischer-H{\"u}bner\inst{4} 
\and Thomas Gro{\ss} \inst{5} 
\and Thomas L{\"a}nger\inst{6} 
\and Mathieu des Noes\inst{7} 
\and Henrich C. P{\"o}hls\inst{8} 
\and Boris Rozenberg\inst{9} 
\and Daniel Slamanig\inst{10}
}
\authorrunning{Thomas Loruenser et al.} 
%
%

\institute
{AIT Austrian Institute of Technology, Vienna \and
ATOS Spain S.A., Spain \and
Technische Universit\"{a}t Darmstadt, Germany \and
Karlstad University, Sweden \and
Newcastle University, UK \and
University of Lausanne, Switzerland \and
Commissariat à l’énergie atomique et aux énergies alternatives, France \and
University of Passau, Germany \and
IBM Haifa Research Lab, Israel  \and
Graz University of Technology, Austria}


\maketitle              

\begin{abstract}
The market for cloud computing can be considered as the major growth area in ICT. However, big companies and public authorities are reluctant to entrust their most sensitive data to external parties for storage and processing. The reason for their hesitation is clear: There exist no satisfactory approaches to adequately protect the data during its lifetime in the cloud. The EU Project \PC (Horizon 2020 programme; duration 2/2015-7/2018) addresses these challenges and yields a portfolio of novel technologies to build security enabled cloud services, guaranteeing the required security with the strongest notion possible, namely by means of cryptography. We present a new approach towards a next generation of security and privacy enabled services to be deployed in only partially trusted cloud infrastructures.



\end{abstract}
\setcounter{footnote}{0}
\input{1_introduction}

\input{2_prismacloud}

\input{3_verifiability}

\input{4_privacy}

\input{5_storage}

\input{6_secandusability}

\input{7_conclusion}

\bibliographystyle{splncs03}
\bibliography{refs}

\end{document}

%% file: 1_introduction.tex
\section{Introduction}

Today, cloud computing is already omnipresent and starts pervading all aspects of our life, whether in the private area or in the business domain. The annual market value related to cloud computing is estimated to be in the region of USD 150 billion, and will probably grow by the year 2018 to around USD 200 billion~\cite{TRA1,PRW1}. 
The European Commission (EC) promotes in its strategy "Digital Agenda for Europe / Europe 2020" the rapid adoption of cloud computing in all sectors of the economy to boost productivity. Furthermore, the EC concludes that “cloud computing has the potential to slash users' IT expenditure and to enable many new services to be developed. Using the cloud, even the smallest firms can reach out to ever larger markets while governments can make their services more attractive and efficient even while reining in spending.”~\cite{STRAT1}.\\[-1.2em]

However, besides these advantages of cloud computing, many new problems arise which are not yet sufficiently solved, especially with respect to information security and privacy. 
The fundamental concept of the cloud is storage and processing by a third party, i.e., the cloud or service provider, which actually invalidates the traditional view of a perimeter in IT security. In fact, the third party becomes part of the company's own computation and storage IT infrastructure albeit not being under its full control. This situation is very problematic and recent incidents show that economic incentives and legal tools used to increase trust in the service provider, e.g. Service Level Agreements, are by far not sufficient to guard personal data and trade secrets against illegal interceptions, insider threats, or vulnerabilities exposing data in the cloud to unauthorized parties. While being processed by a provider, data is typically neither adequately protected against unauthorized read access, nor against unwanted modification, or loss of authenticity. Consequently, in the most prominent cloud deployment model today -- the public cloud -- the cloud service provider necessarily needs to be trusted. Security guarantees with respect to user data can only be given on a contractual basis and rest to a considerable extent on organisational (besides technical) precautions. Hence, outsourcing IT tasks to an external shared infrastructure builds upon a problematic trust model. This situation inhibits many companies in the high-assurance and high-security area to benefit from external cloud offerings: for them confidentiality, integrity, and availability are of such major importance that adequate technical measures are required---but state-of-the-art ICT can currently not provide them. Moreover, individuals using public cloud services face a considerable privacy threat too, since they typically expose more information to services than required to perform the task. In all cases, in the end the cloud user is responsible for his or her data and outsourcing sensitive tasks to an external entity does not remove this burden. Therefore, novel security and privacy preserving methods need to be developed to facilitate cloud usage even for organisations dealing with sensitive information.

In this work we present a new approach towards cloud security which is developed by the \PC consortium within the EU Horizon 2020 research framework. The vision of \PC is to develop the next-generation of cryptographically secured cloud services with security and privacy built in by design. For us, the only reasonable way to achieve the required security properties for outsourced data storage and processing is by adopting suitable cryptographic mechanisms. \PC shall impact through the development of next generation secure cloud services $(i)$~to achieve beneficial impact in society, industry, and research in Europe
$(ii)$~to remove a major inhibitor against cloud adoption in security relevant domains $(iii)$~by developing cloud applications, that preserve more privacy for citizens,
$(iv)$~for delivering input and strengthening the position of European industries, 
$(v)$~to strengthen European research in a field with high research competition.


%% file: 2_prismacloud.tex
\section{A New Take on Cloud Security}
\label{sec:goals}

\subsection{Relevance and Project Objectives}

The importance of security for cloud computing is now widely accepted \cite{CSA2,ENISA1,NIST1} and security research for cloud computing is gaining tremendous momentum. It is of major importance for security research to catch up with the rapid developments in cloud computing before the ongoing transition to the cloud reaches critical areas. Future solutions for cloud deployment must be secure by design and provide end-to-end security for users to the best extent possible.

In the long-term, cloud computing will influence many important aspects of our lives and will likely have an enormous impact on our society. The European Commission already recognised the potential future impact of cloud computing for all of us and has issued a cloud computing strategy \cite{STRAT1} to protect European citizens from potential threats, while simultaneously unleashing the potential of cloud computing, for both the industry/public sector as well as for individuals. Many initiatives with these goals are currently either ongoing, or are going to be formed in the nearer future, and a substantial effort is put into cloud security research through the EU-research framework. However, there are still many unsolved problems and there is ample room for innovation. We will contribute innovations to that field and are targeting the development of next-generation secure and trustworthy cloud environments. The ambition of \PC is to build trustworthy cloud services on top of untrusted infrastructure by the development and application of adequate cryptography and methodologies.

The main objectives of \PC are: $(i)$~to develop next-generation cryptographically secured services for the cloud. This includes the development of novel cryptographic tools, mechanisms, and techniques
ready to be used in a cloud environment to protect the security of data over its lifecycle and to protect the privacy of the users. The security shall be based on 'by design' principles. $(ii)$~to assess and validate the project results by fully developing and implementing three realistic use case scenarios in the areas of e-government, healthcare, and smart city services. $(iii)$~to conduct a thorough analysis of the security of the final systems, their usability, as well as legal and information governance aspects of the new services. 

\subsection{State of the Art}

Ongoing research activities like SECCRIT, Cumulus, and PASSIVE\footnote{EU-FP7: \url{http://www.seccrit.eu/}, \url{http://www.cumulus-project.eu/}, \url{http://ict-passive.eu/}} are extremely valuable and will be setting the standards and guidelines for secure cloud computing in the next years. However, these approaches consider the cloud infrastructure provider as being trustworthy in the sense that no information of the customers, i.e., tenants, will be leaked, nor their data will be tampered with. The cloud infrastructure provider, however, has unrestricted access to all physical and virtual resources and thus absolute control over all tenants' data and resources. The underlying assumption is, that if the cloud  provider performs malicious actions against its customers, in the long run, he or she will be put out of business -- if such doings are revealed. However, this assumption is very strong, especially considering the ongoing revelation of intelligence agencies' data gathering activities. Data disclosure may even be legally enforced in a way completely undetectable by the cloud provider's customers.

Through auditing and monitoring of cloud services, some of the malicious behaviour of outsiders and insiders (e.g. disgruntled employees with administrator privileges) may be detectable \emph{ex-post}, however, that does not help a specific victim to prevent or survive such an attack. Moreover, advanced cyber-attacks directly targeting a specific victim can barely be detected and prevented with cloud auditing mechanisms or anomaly detection solutions. These methods are more efficient for the detection of large scale threats and problems and for making the infrastructure itself resilient, while keeping an acceptable level of service.

Other projects, like TClouds and PRACTICE\footnote{EU-FP7: \url{http://www.tclouds-project.eu}, \url{http://www.practice-project.eu/}} take cloud security a step further: TClouds already considers the impact of malicious provider behaviour and tries to protect users. However, it is not strongly focusing on comprehensive integration of cryptography up to the level of end-to-end security. PRACTICE, in contrast, is well aligned with our idea of secure services. However, it focuses mainly on the preservation of data confidentiality for processing, when outsourced to the cloud. This is achieved by means of secure multiparty computations and concepts from fully homomorphic encryption. \PC is complimentary to these concepts and enhance them with cryptographic primitives for the verification of outsourced computation and other relevant functionality to be carried out on the data in the untrusted cloud.

Research activities in context of privacy in cloud computing were and are currently conducted by various projects like ABC4Trust, A4Cloud and AU2EU\footnote{EU-FP7: \url{https://abc4trust.eu}, \url{http://www.a4cloud.eu}, \url{http://www.au2eu.eu}}. \PC complements these efforts by relying on and further developing privacy-enhancing technologies for the use in cloud based environments.

\subsection{Main Goal and Innovations}

The main goal of \PC is to enable the deployment of highly critical data to the cloud. The required security levels for such a move shall be achieved by means of novel security enabled cloud services,  pushing the boundary of cryptographic data protection in the cloud further ahead. \PC core innovations include:
$(i)$~Techniques for outsourcing computation with verifiable correctness and authenticity-preservation for allowing secure delegation of computations to cloud providers. 
$(ii)$~The provision of cryptographic techniques for the verification of claims about the secure connection and configuration of the virtualized cloud infrastructures. 
$(iii)$~Addressing user privacy issues by cryptographic data minimization and anonymization technologies. 
$(iv)$~Improving anonymization techniques for very large data sets in terms of performance and utility.
$(v)$~A distributed multi-cloud data storage architecture for sharing data among several cloud providers and thus improving data security and availability. Dynamically updating distributed data by means of novel techniques shall avoid vendor lock-in and promote a dynamic cloud provider market, while preserving data authenticity and facilitating long term data privacy with proactive secret sharing.  
$(vi)$~Advanced format and order preserving encryption and tokenisation schemes for seamless integration of encryption into existing cloud services.
$(vii)$~The \PC work program is complemented with activities addressing secure user interfaces, secure service composition, secure implementation in software and hardware, security certification, and an impact analysis from an end-user view. In order to converge with the European Cloud Computing Strategy, a strategy for the dissemination of results into standards is developed. 
$(viii)$~As feasibility proof, three use cases from the fields of SmartCity, e-Government, and e-Health will be fully implemented and evaluated by the project participants.

%% file: 3_verifiability.tex
\section{Verifiability of Data, Processing and Infrastructure}
\label{sec:verifiablity}

\subsection{Verifiable and Authenticity Preserving Data Processing}

Verifiable computing aims at outsourcing computations to one or more untrusted processing units in a way that the result of a computation can be efficiently checked for validity. General purpose constructions for verifiable computations have made significant process over the last years \cite{DBLP:journals/cacm/WalfishB15}, there are already various implemented systems which can be deemed nearly practical, but are not yet ready for real-world deployment. Besides general purpose systems, there are other approaches that are optimized for specific (limited) classes of computations or particular settings, e.g, \cite{DBLP:conf/ccs/BackesFR13,DBLP:conf/ccs/FioreGP14,DBLP:conf/asiacrypt/CatalanoMP14}. 

In addition to verifiability of computations, another interesting aspect is to preserve the authenticity of data that is manipulated by computations. Tools for preserving authenticity under admissible modifications  are (fully) homomorphic signatures (or message authentication codes) \cite{DBLP:conf/scn/Catalano14}. 
Besides this general tool, there are signatures with more restricted capabilities, like redactable signatures introduced in~\cite{johnson2002_RSS,Steinfeld2002_RSS}, which have recently shown to offer interesting applications~\cite{poehlsACNS2014updatableRSS,Hanser2013BlankDigitalSignatures}. These and other functional and malleable signatures will be developed further within \PC to meet requirements set by cloud applications. 

By combining these cryptographic concepts, \PC aims at providing tools that allow to realize processes (with potentially various participating entities) that guarantee to preserve the authenticity and provide verifiable of involved data and computations respectively.    

%
%
%

\subsection{Integrity and Certification of Virtualized Infrastructure}

The area of structural integrity and certification of virtualized infrastructures bridges between three areas: attestation of component integrity, security assurance of cloud topologies and graph signatures \cite{Gros2014,Gros2015} to connect these areas. Attestation is the process in which a trusted component asserts the state of a physical or virtual component of the virtualized infrastructure, on all the layers of it.
Cloud security assurance refers to a research area and line of cloud security tools, in which recent proposals included the analysis of cloud topologies for security properties.
Graph signatures, that is, signatures on committed graphs are a new primitive we investigate within \PC. Such a signature scheme allows a recipient of a signature to commit a hidden graph and have an issuer sign the recipient-committed graph while joining it with an issuer-committed graph. The resulting signature allows the recipient to prove in zero-knowledge properties of the graph, such as connectivity isolation. 

Within \PC we develop and optimize the use of graph signatures for practical use in virtualized infrastructures. Their application allows an auditor to analyse the configuration of a cloud, and issue a signature on its topology. The signature encodes the topology as a graph in a special way, such that the cloud provider can use it to prove in zero-knowledge high-level security properties such as isolation of tenants to verifiers, such as the tenants, without disclosure of secret information. 
Further, we will bride between cloud security assurance and verification methodology and certification by establishing a framework that issues signatures and proves security properties based on standard graph models of cloud topologies and security goals in formal language (VALID).

%% file: 4_privacy.tex
\section{User Privacy Protection and Usability}
\label{sec:privacy}

\subsection{Privacy Preserving Service Usage}

For many services in the cloud it is important that users are given means to prove their authorisation to perform or delegate a certain task. However, it is not always necessary that users reveal their full identity to the cloud, but only prove by some means that they are authorised, e.g., possess certain rights. The main obstacle in this context is, that a cloud provider must still be cryptographically reassured that the user is authorised. 

Attribute-based anonymous credential (ABC) systems have proved to be an important concept for privacy-preserving applications as they allow users to authenticate in an anonymous way without revealing more information than absolutely necessary to be authenticated at a service and there are strong efforts to bring them to practice\footnote{e.g., ABC4Trust: \url{https://abc4trust.eu/}}. Well known ABC systems are for instance the multi-show system Idemix \cite{DBLP:conf/ccs/CamenischH02} and the one-show system U-Prove \cite{uprove-spec}. Recently also some alternative approaches for ABC systems from malleable signature schemes \cite{Canard2013,DBLP:conf/csfw/ChaseKLM14} and a variant of structure-preserving signatures \cite{DBLP:conf/asiacrypt/HanserS14} have been proposed.

In \PC we aim at improving the state of the art in anonymous credential systems and group signature schemes with a focus on their application in cloud computing services. Besides traditional application such as for anonymous authentication and authorization we will also investigate their application to privacy-preserving billing \cite{DBLP:conf/ih/DanezisKR11,Sl11} for cloud storage and computing services. 

\subsection{Big Data Anonymization} 

Anonymizing data sets is a problem which is often encountered when providing data for processing in cloud applications in a way, that a certain degree of privacy is guaranteed. However, e.g. achieving optimal $k$-anonymity is known to be an NP-hard problem. Typically, researchers have focused on achieving $k$-anonymity with minimum data loss, thus maximizing the utility of the anonymised results. But all of these techniques assume that the dataset to be anonymised is relatively small (and fits into computer memory). In the last few years several attempts have been made to tackle the problem of anonymising large datasets.

In \PC, we aim to improve existing anonymisation techniques in terms of both  performance and utility (minimizing information loss) for very large data sets. We strive to overcome deficiencies in current mechanisms, e.g. size limitations, speed, assumptions about quasi-identifiers, or existence of total ordering, and implement a solution suitable for very large data sets. In addition, we propose to address issues related to distribution of very large data sets.

%% file: 5_storage.tex
\section{Securing Data at Rest}
\label{sec:storage}

\subsection{Confidentiality and Integrity for Unstructured Data} 

Protecting customer data managed in the cloud from unauthorised access by the cloud provider itself should be one of the most basic and essential functionalities of a cloud system. However, the vast majority of current cloud offerings does not provide such a functionality. One reason for this situation is, that current cryptographic solutions can not be easily integrated without drastically limiting the capabilities of the storage service. 

In PRISMACLOUD, we aim at researching and developing  novel secure storage solutions with increased flexibility based on secret sharing. Secret sharing was invented in the late nineteen-seventies and has become a vital primitive in many cryptographic tasks. It can also be used to provide confidentiality for data at rest with strong security in a key-less manner when working in a distributed setting. Various systems have been proposed during the last years, but most of them work in rather naive single user approaches and require a trusted proxy in their setting. First approaches to support multiple users have been proposed in a combination with quorum based meta-data management, but still rely on passive storage nodes. Recently, a new type was proposed, using active nodes to fully delegate secure multi-user storage to the cloud. It combines efficient Byzantine protocols with various types of secret sharing protocols to cope with different adversary settings in a flexible way. However, many desired features, as well as a trustworthy distributed access control mechanism are still missing. 

Our goal is to develop efficient and flexible secret sharing based storage solutions for dynamic environments, like the cloud, supporting different adversary models (active, passive, mixed). The research will focus on the design of multi-user storage systems in a distributed fashion, without single-point-of-trust and single-point-of-failure and how they can be extended with access privacy. 



\subsection{Long-term Security Aspects and Everlasting Privacy}

To provide protection goals, such as integrity, authenticity, and confidentiality in the long-term, classic cryptographic primitives like digital signatures and encryption schemes are not sufficient.
They become insecure when their
security properties are defeated by advances in computer power or cryptanalytic techniques.
Thus, the only approach known to address long-term confidentiality is by using proactive secret sharing, e.g. \cite{Gupta:2007:GVI:1338446.1338890}.
In this approach, the data is split into several shares that are stored in different locations and are renewed from time to time.
Although secret sharing is needed to provide long-term confidentiality, there is no approach that considers performing publicly or privately verifiable computations or integrity preserving modifications on secret shares yet.
Besides the distributed storage of data, to provide everlasting privacy (or confidentiality) for data processed in a publicly verifiable manner, the information published for auditing needs to be information-theoretically secure.
Only a few solutions address this and only for specific problems, such as verifiable anonymisation of data \cite{DBLP:conf/fc/BuchmannDG13} and verifiable tallying of votes, e.g. \cite{DBLP:journals/tissec/MoranN10}.
No general applicable solution is provided, nor do existing approaches show how authenticated data can be processed in a publicly verifiable way.
Therefore, we aim at providing solutions for proactive secret sharing of authenticated data and techniques that allow for privately and publicly verifiable computations.

\subsection{Cryptography for Seamless Service Integration}

Considering existing applications in the cloud, it may be impossible to transparently add security features later on, e.g., to store encrypted data in the same database table used for unencrypted data, and applications running on the database may be unable to use the encrypted data, causing them to crash or alternatively, to output incorrect values. Standard encryption schemes are designed for bit-strings of a fixed length, and can therefore significantly alter the data format, which may cause disruptions both in storing and using the data.


To address this problem, techniques like Format-Preserving Encryption (FPE), Order-Preserving Encryption (OPE) and Tokenizaiton have emerged as most useful tools. In FPE schemes the encrypted ciphertexts have the same format as the messages, i.e. they can be directly applied without adapting the application itself. OPE schemes on the other hand, maintain the order between messages in the original domain, thus allowing execution of range queries on encrypted data. 

In \PC we aim to address the shortcomings of the existing FPE and OPE schemes. It can be shown that existing FPE schemes for general formats, e.g. name, address, etc., are inefficient, lack in their security level, and do not provide a clear way for format definition, thus making them practically unusable.
We propose to address both issues (security and efficiency) and develop an FPE scheme for general formats that: $(i)$ is more efficient; $(ii)$ provides an acceptable security guarantee; $(iii)$ supports complex format definition; $(iv)$ could be employed to solve practical problems, e.g. data sharing for cluster of private clouds. For OPE we aim to analyze further the existing approaches from both security and performance perspectives (and/or develop our own technique) and implement the option that provides a suitable security-functionality trade-off for a given set of applications/use cases.

%% file: 6_secandusability.tex
\section{Methodology, Tools and Guidelines for Fast Adoption}
\label{sec:sechci}

\subsection{Holistic Security Models}
The paradigm of service orientation~\cite{Erl06} has increasingly been adopted as one of the main approaches for developing complex distributed systems out of re-usable components called services. We want to use the potential benefits of this software engineering approach, but not build yet another semi-automated or automated technique for service composition. However, combining the building blocks of \PC correctly would require the developers to have a solid understanding of their cryptographic strength.  

To allow composing the building blocks into secure higher level services, we will identify which existing models for the security of compositions are adequate to deal with the complexity and heterogeneity. 
\PC will adopt working and established solutions and assumes that the working way of composing services can be a way to allow secure composition. When each service can be described using standard description languages this allows extending composition languages~\cite{BBG06} to provide further capabilities, e.g., orchestrations, security, transaction, to service-oriented solutions~\cite{PLS08}. In \PC we want to reduce the complexity further, just like recently, mashups~\cite{LHPB09} of web APIs provided means for non-experts to define simple workflows.
Within \PC we will develop a description of not only the functionality of each cryptographic building block but also of their limitations and composability.

\subsection{Human Computer Interaction (HCI) Concepts}

Cryptographic concepts such as secret sharing, verifiable computation or anonymous credentials, are fundamental technologies for secure cloud services and to preserve end users' privacy by enforcing data minimization. End users are still unfamiliar with such data minimization technologies that are counterintuitive to them and for which no obvious real-world analogies exist. In previous HCI studies, it has been shown that users have therefore difficulties to develop the correct mental models for data minimisation techniques such as anonymous credentials \cite{DBLP:conf/ifip11-4/WastlundAF11} or the new German identity card \cite{harbach2013acceptance}.  Moreover, end users often do not trust the claim that such privacy-enhancing technologies will really protect their privacy \cite{andersson2005trust}. Similarly, users may not trust claims of authenticity and verifiability functionality of malleable and of functional signature schemes.
In our earlier research work, we have explored different ways in which comprehensive mental models of the data minimization property of anonymous credentials can be evoked on end users \cite{DBLP:conf/ifip11-4/WastlundAF11}. \PC extends this work by conducting research on suitable metaphors for evoking correct mental models for other privacy-enhancing protocols and cryptographic schemes used in \PC.
Besides, it researches what social trust factors can establish trust in \PC technology and how this can be matched into the user interfaces. 

\subsection{Secure Cloud Usage for End-Users}

The cloud computing services market is currently soaring and already in the range of USD 100 billion, with an outlook onto a bright economic future with projected steep annual growth rates varying between 10 and 20\% \cite{TRA1,PRW1}. The huge interest in influencing market development and securing a proper share of the market is also visible in the manifold industry-driven standardization efforts\footnote{several consortia are listed at \url{http://cloud-standards.org}}. Following \emph{Key Action 1 ``Cutting through the Jungle of Standards''} of the European Commission's Cloud Computing Strategy \cite{STRAT1}, we will take a cautious approach with regard to a potential introduction of project results into a standardization process.

The crucial role of data security in cloud applications is widely recognized. 
Previous studies have shown the vulnerability of information and communication technology systems, and especially also of cloud systems, to illegal and criminal activities \cite{SGH1}. We will take a critical appraisal of the secure cloud systems proposed in \PC and will analyze, whether they live up to the security promises in practical application. We will give an indication for individuals, and for corporate and institutional security managers, what it means in practice to entrust sensitive data in specific use cases to systems claiming to implement, e.g.,``everlasting privacy'' \cite{MQU1}. 

Besides licit use, we will assess the impact of potential criminal uses and misuses of the secure cloud infrastructures to foster, enhance, and promote cybercrime. We want to anticipate threats resulting from misuse, deception, hijacking, or misappropriation by licit entities.
We will map implications originating in technical details and in the operation or usage of systems in specific environments, to high level security objectives which can be understood and relied on in high-level security management practice \cite{LAE1}. 

%% file: 7_conclusion.tex
\section{Conclusion}
\label{sec:conclusion}

According to the importance of the project goals, i.e., to enable secure dependable cloud solutions, \PC will have a significant impact in many areas. On a European level, \PC's disruptive potential of results lies in its provision of a basis for the actual implementation and deployment of security enabled cloud services. Jointly developed by European scientists and industrial experts, the technology can act as an enabling technology in many sectors, like health care, e-government, smart cities. Increasing adoption of cloud services, with all its positive impact on productivity, and creation of jobs may be stimulated. On a societal level, \PC potentially removes a major roadblock towards the adoption of efficient cloud solutions to a potential benefit of the end-users. Through the use of privacy-preserving data minimization functionalities, and depersonalization features, the amount of data being collected about end-users may effectively be reduced, maintaining the full functionality of the services. We will explicitly analyse potential negative consequences, and potential misuses (cybercrime) of secure cloud services. The potential impact for European industry is huge: \PC results may contribute to pull some of the business currently concentrated in the United States of America to Europe and create sustainable business opportunities for companies in Europe. Equally important is the potential impact of \PC for the European scientific community, as its results will be very much on the edge of scientific research.